\documentclass[aps,prb,twocolumn,showpacs,groupedaddress,email]{revtex4-1}  

\usepackage{graphicx}  
\usepackage{bm}        
\usepackage{amssymb}   
\usepackage{amsmath}

\newcommand{\bra}[1]{\left\langle #1\right|}
\newcommand{\ket}[1]{\left| #1\right\rangle}

\newcommand{\mean}[1]{\langle #1\rangle}
\renewcommand{\imath}{{\rm i}}

\begin{document}
\title{Large suppression of quantum fluctuations\\ of light from a single emitter by an optical nanostructure}

\author{Diego Mart\'in-Cano,$^{1,3}$ Harald R. Haakh,$^{1}$
Karim Murr$^{2,3,4,5}$ and Mario Agio$^{2,3,4}$}
\affiliation{
$^{1}$Max Planck Institute for the Science of Light, 91058 Erlangen, Germany\\
$^{2}$National Institute of Optics (CNR-INO), 50125 Florence, Italy\\
$^{3}$Center for Quantum Science and Technology in Arcetri (QSTAR), 50125 Florence, Italy\\
$^{4}$European Laboratory for Nonlinear Spectroscopy (LENS), 50019 Sesto Fiorentino, Italy\\
$^{5}$Dipartimento di Fisica ed Astronomia, Universit\`a di Firenze, 50019 Sesto Fiorentino, Italy
}

\begin{abstract}
We investigate the reduction of the electromagnetic field fluctuations in resonance fluorescence from a single emitter coupled to an optical nanostructure.
We find that such hybrid system can lead to the creation of squeezed states of light, with quantum fluctuations significantly below the shot noise level. Moreover, the physical conditions for achieving squeezing are strongly relaxed with respect to an emitter in free space. A high degree of control over squeezed light is feasible both in the far and near fields, opening the pathway to its manipulation and applications on the nanoscale with state-of-the-art setups.
\end{abstract}

\maketitle

\section{Introduction}
Optical nanostructures are known to be
efficient architectures for controlling light-matter interactions.\cite{novotny11}
In this context, the most widely considered processes have been 
Raman scattering and fluorescence, whose enhancement has been experimentally verified at the single-emitter level.\cite{kneipp97,anger06,kuehn06a}
A major goal is now to explore their performance in
the quantum regime,\cite{tame13} so far mainly examined in cavity
quantum electrodynamics.\cite{raimond06}
Antibunching has been investigated as a signature of the granularity of quantum light arising
from single emitters coupled to nanostructures.\cite{akimov07,schietinger09,huck11}
In contrast, electromagnetic field fluctuations below shot noise,\cite{walls81} which mirror the quantum wave nature of light, are known to be challenging to measure\cite{ourjoumtsev11} at the quantum level and
have not been addressed in such hybrid systems.

Reduced quantum fluctuations 
are the unique characteristics of squeezed states of light,\cite{walls83} which are relevant for overcoming classical application limits in, for instance, precision measurements, spectroscopy and optical communications.
Despite recent advances on the microscopic
scale,\cite{ourjoumtsev11,safavi-naeini13}
sources of squeezed light usually rely on the nonlinear
response of macroscopic systems, typically crystals or atomic
vapors.\cite{loudon87}
Although optical nanostructures exhibit classical field statistics
in the linear regime, they are able to fundamentally alter
the radiation properties of a quantum emitter (QE) placed at close
proximity.\cite{novotny11}
This approach can be applied in a broad range of nanoarchitectures and QEs,
covering atoms,\cite{stehle11} color centers,\cite{schietinger09,huck11} molecules,\cite{anger06,kuehn06a} or quantum dots.\cite{akimov07}
An interesting question is thus to what extent the coupling between
a nanostructure and a QE can modify
the electromagnetic field fluctuations.

Here, we show that nanostructures can significantly increase squeezing
in the resonance fluorescence from a QE.
Moreover, they strongly relax the conditions for overcoming shot noise
in terms of bandwidth and excitation power.
Our results open a pathway towards the experimental measurement of such
squeezed states of light in state-of-the art setups and their manipulation
on the nanoscale, with prospects for advancing applications at the
single-photon level.

\section{Discussion}
\paragraph*{Method.}
Electromagnetic field fluctuations can be measured by homodyne
techniques,\cite{vogel06} which detect the variance
of the electric field quadrature component
${\hat{E}_{i}(\mathbf{r},t)=\hat{E}^{(+)}_{i}(\mathbf{r},t)+\hat{E}^{(-)}_{i}(\mathbf{r},t)}$, given by ${(\Delta \hat{\mathcal{E}_{i}})^{2}=\mean{:\!(\hat{E_{i}}-\mean{\hat{E_{i}}})^{2}\!:}}$.
Here, we consider the normal ordering ($: :$)
to directly compare the variance to the
shot-noise level, so that negative values of $(\Delta \hat{\mathcal{E}_{i}})^{2}$
indicate squeezed light.
We evaluate these fluctuations in the framework of macroscopic quantum electrodynamics in dispersive and absorptive media.\cite{vogel06}
In the case of a two-level QE and imposing the rotating wave and Markov approximations, the positive-frequency scattered electric field operator is
$
\label{Eq1Efield}
\hat{E}_{i}^{(+)}(\mathbf{r},t)=|g_i(\mathbf{r})
|e^{\imath\phi_i(\mathbf{r})}\mathbf{\hat{\sigma}}(t)\,,
$ 
which depends on the QE coherence $ {\hat{\sigma}={\ket{g}\!\bra{e}}}$. Here, $\ket{g}$ and $\ket{e}$ are the QE's ground and excited states,
respectively.
The emission characteristics in the presence of a given nanoarchitecture are encoded in the amplitude $|g_i|$ and phase $\phi_i$, which can be expressed in terms of the classical electromagnetic Green's tensor\citep{Tai96} (see details in Appendix~\ref{A1}).
Evaluating the fluctuations of $\hat{E}_{i}(\mathbf{r},t)$ we find
\begin{widetext}
\begin{equation}
\label{Eq4Efieldvariance}
{(\Delta\hat{\mathcal{E}}_{i}(\mathbf{r},t))^{2}}
 =
2|g_i(\mathbf{r})|^2
\biggl[\left(
\mean{\mathbf{\hat{\sigma}}^{\dag}(t)\mathbf{\hat{\sigma}}(t)}
-|\mean{\mathbf{\hat{\sigma}}(t)}|^2\right)
 -
\textrm{Re}\left(e^{\imath 2\phi_i(\mathbf{r})}
\mean{\mathbf{\hat{\sigma}}(t)}^{2}\right)\biggr].
\end{equation}
\end{widetext}
The expectation values are then replaced by the solution of the optical Bloch equations
  under steady-state conditions.
These contain the effects of the driving field's Rabi frequency $\Omega$, the spontaneous decay at a rate $\gamma$, and the frequency detuning $\delta_\mathrm{L}=\omega_\mathrm{L}-\tilde{\omega}_\mathrm{E}$ between the laser and the QE. The nanostructure affects all of these via 
a local field enhancement and a shift in the QE resonance to a value $\tilde \omega_E$.
We also allow for additional pure dephasing at a rate $\gamma^*$.\cite{vogel06}
In this case, Eq.~\eqref{Eq4Efieldvariance} can be expressed in a form valid for a QE in any environment
\begin{widetext}
\begin{equation}
\label{Eq5variancedephasing}
{(\Delta\hat{\mathcal{E}}_{i}(\mathbf{r},t))^{2}}
 \underset{\rm steady~state}{=}
|g_i(\mathbf{r})|^2\frac{z^2}{1+\delta^2+z^2}
\left(1-
\frac{(\delta^2+1)
(1+\cos[2\phi_{i}+2\Phi-2\omega_\mathrm{L}t])}
{(1+x)(1+\delta^2+z^2)}\right),
\end{equation}
\end{widetext}
expressed in terms of the normalized dephasing rate $x=2\gamma^{*}/\gamma$,
the normalized detuning $\delta=2\delta_\mathrm{L}/(\gamma+2\gamma^{*})$,
and the normalized Rabi frequency
$z=\sqrt{2}|\Omega|/\sqrt{\gamma (\gamma+2\gamma^{*})}$, associated with the QE's saturation parameter.
The cosine in Eq.~(\ref{Eq5variancedephasing}) can be set to unity without loss of generality.\footnote{The cosine in Eq. (\ref{Eq5variancedephasing}) contains phases
originating from scattering off the nanostructure and from the evolution of
the coherence operator, and can always be set to unity, for instance by choosing an appropriate homodyne scheme for the detection of squeezing
(see Appendix~\ref{A3})
Therefore, without loss of generality, we impose this in the numerical evaluations.}
A detailed derivation of Eqs. (\ref{Eq4Efieldvariance}) and (\ref{Eq5variancedephasing}) are given in Appendix~\ref{A2}.

\paragraph*{A hybrid nanosystem.}
From Eq.~\eqref{Eq4Efieldvariance}, we see that the electric field
fluctuations generated by a QE in resonance fluorescence are governed by the emitter's
optical coherence $\mathbf{\hat{\sigma}}$
and
upper-state population $\hat{\sigma}^{\dag}\hat{\sigma}$.\cite{walls81}
The fluctuations
$\mean{\mathbf{\hat{\sigma}}^{\dag}(t)\mathbf{\hat{\sigma}}(t)}-
|\mean{\mathbf{\hat{\sigma}}(t)}|^2$ are always positive,
and, hence, tend to destroy squeezing, but they approach zero the weaker the excitation.
Since we deal with one QE, this results in low photon count rates,
which has prevented the detection of fluctuations below shot noise
in free space and made it challenging even in the presence of a resonator.\cite{ourjoumtsev11}
The last term in Eq.~(\ref{Eq4Efieldvariance}) originates from
quantum fluctuations in the optical coherence. It is the only one
able to create squeezing and cannot be interpreted neither with
classical waves nor with particles alone, bearing out the quantum wave
nature of this process.
%
%
If a QE is placed near a nanostructure, the dynamics that generate
quantum squeezing are fundamentally changed.
First, both the amplitude $|g_i(\mathbf{r})|^2$ and the phase $\phi_i(\mathbf{r})$
of the field fluctuations are modified by the nanostructure
due to its electromagnetic response.
Hence, although the nanostructure increases the field intensity
scattered by the QE.\cite{novotny11}
its quantum fluctuations can be comparatively
reduced with respect to shot noise, with a squeezing amplitude $|g_i(\mathbf{r})|^2$.
Second,
since the coherence $\hat{\sigma}$ is affected by
the enhancement of the driving field and the change in the
spontaneous decay rate, both induced by the nanostructure,\cite{novotny11}
control of these magnitudes can be used to significantly reduce the
electromagnetic field fluctuations in the emission from a QE,
while increasing the photon count rate.
%

%
\begin{figure}[!h]
\begin{center}
\includegraphics[width=8.25cm,angle=0]{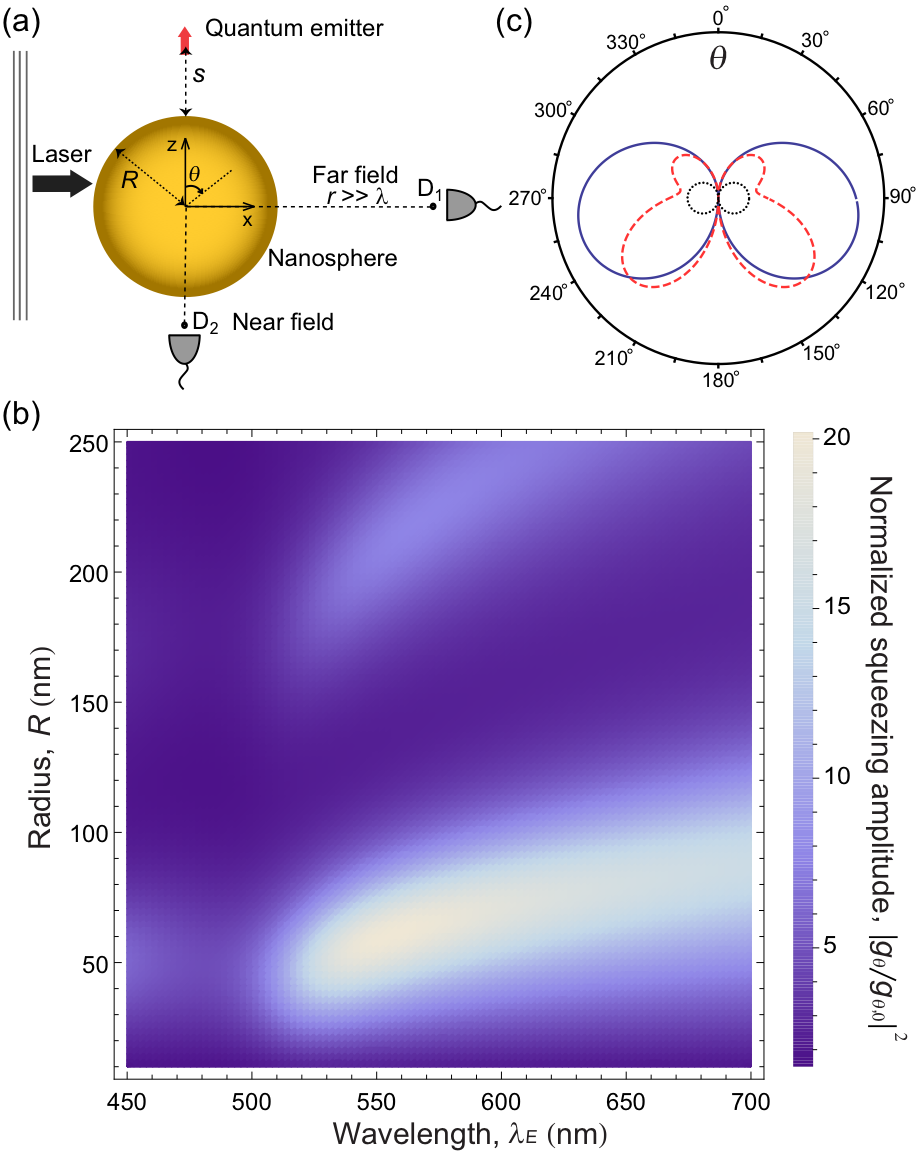}
\end{center}
\caption{(Color online) 
(a) Hybrid system consisting of a quantum emitter at a distance $s$ from a gold nanosphere of radius $R$.
$D_1$ and $D_2$ are the detection points in the far
and near fields, respectively.
$D_1$ is on the $x$-axis at
a distance $10^{5}\lambda_\mathrm{E}$ from the nanosphere center,
while $D_2$ is along the $z$-axis, 10 nm from the nanosphere surface.
The emitter dipole moment is oriented perpendicularly to the
nanosphere surface.
(b) Normalized squeezing amplitude $|g_{\theta}/g_{\theta,0}|^{2}$
as a function of $\lambda_\mathrm{E}$ and $R$.
For comparison, the squeezing amplitude is normalized with respect to its value  in the absence of the nanosphere, $|g_{\theta,0}|^2$.
The distance between the quantum emitter and the nanosphere surface is $s=10$ nm.
The detection point corresponds to $D_1$ as shown in panel (a).
The $\theta$-component of the field quadrature corresponds
to the dominant polarization in this configuration.
(c) Far-field squeezing amplitude $|g_{\theta}|$, near the dipolar
($R=60$ nm, solid blue curve)
and quadrupolar ($R=120$ nm, dashed red curve)
nanosphere resonances at $\lambda_\mathrm{E}=550$ nm.
The dotted black curve corresponds to the free-space case.\label{fig1}}
\end{figure}

\paragraph*{Far-field squeezing amplitudes.}
For a quantitative analysis, we exemplify the nanostructure with a gold nanosphere (GNS), coupled to a QE characterized by its transition frequency $\omega_E = 2 \pi c / \lambda_E$ and dipole $\boldsymbol{d}$,
as illustrated in Fig.~\ref{fig1}a.
In this case, the Green's tensor $\boldsymbol{G}$ is known analytically.
In the far field ($|\mathbf{r} - \mathbf{r}_{E}|\gg \lambda_E$),
 $g_i = |g_i|e^{\imath \phi} \approx  \frac{\omega_\mathrm{E}^{2}}{\varepsilon_{0}c^{2}} G_{ij}(\mathbf{r},\mathbf{r}_\mathrm{E},\omega_\mathrm{E})d_{j}$ provides an excellent approximation of the amplitude and phase in Eq.~\eqref{Eq4Efieldvariance}, whereas
a quantum correction must be included in the near field\cite{dung02}
(see further details in Appendix~\ref{A1}).
Figure~\ref{fig1}b shows the squeezing amplitude $|g_{\theta}|^2$ at the detection point in the far field ($D_1$ in Fig.~\ref{fig1}a),
where the $\theta$-component dominates.
This amplitude features several local maxima, arising from the excitation of
plasmon-polariton resonances,\cite{bohren83b}
which depend on the nanosphere radius $R$
and on the QE emission wavelength $\lambda_\mathrm{E}$.
The strongest one originates from the dipole resonance, as
indicated by the two-lobe pattern
in the far field, shown in Fig.~\ref{fig1}c.
Notice that near the global maximum, squeezing is enhanced by a factor of 20 due to the presence of the nanosphere.
Further maxima at larger radii are associated with higher-order resonances.
Although they provide weaker enhancement,
they reshape the far-field pattern more strongly than the dipolar one
(see Fig.~\ref{fig1}c).
Therefore, nanostructures may be exploited to control the
directionality of squeezed-state emission in the far field,
which can be optimized by suitably designed
architectures.\cite{curto10}

\begin{figure}[!h]
\begin{center}
\includegraphics[width=8.25cm,angle=0]{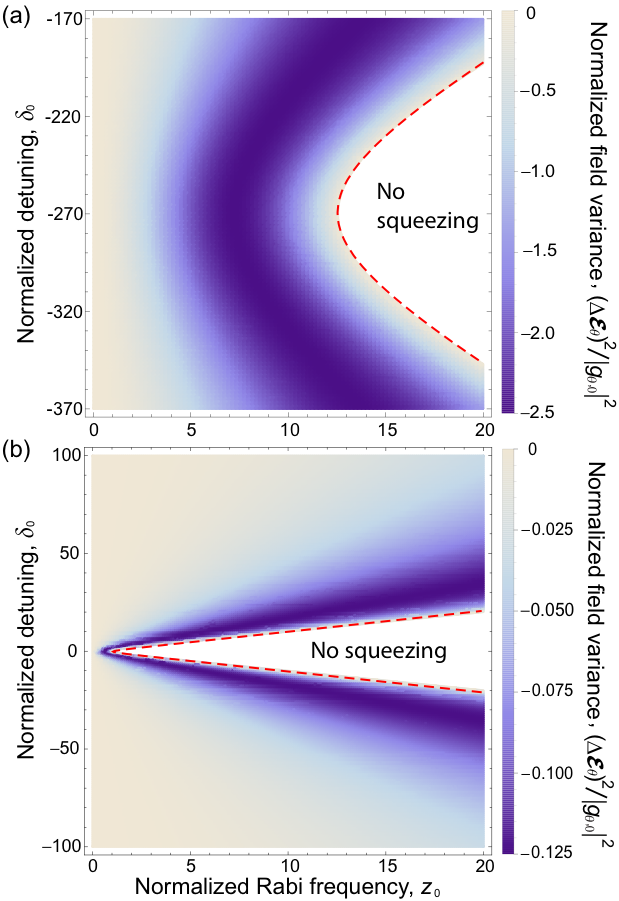}
\end{center}
\caption{(Color online) 
Electric field fluctuations in the presence (a) and in the absence of a nanosphere (b).
%
The relevant system parameters are $s=10$ nm, $R=60$ nm, $\lambda_\mathrm{E}=550$ nm and fields are detected at $D_1$ (see Fig.~\ref{fig1}a).
For comparison, the variances are normalized by the squeezing amplitude in free space $|g_{\theta,0}|^{2}$.
Both panels cover equal ranges of detuning and driving laser intensity.
Moreover, the lower bound of the color scale displays the different minimum value in each panel.
Their ratio emphasizes the 20-fold enhancement of squeezing due to the nanosphere  as compared to free space.
\label{fig2}
}
\end{figure}

\paragraph*{Bounds of squeezing.}
The presence of the nanostructure also strongly modifies the conditions
for the creation of squeezed light from a QE.
This is possible because the field fluctuations depend on the frequency detuning
$\delta_\mathrm{L}$ between the QE and the driving field,
the Rabi frequency $\Omega$ (\emph{i.e.} the
driving field) and the QE's spontaneous decay rate $\gamma$,
which differ from their values in free space\cite{novotny11} ($\delta_{\mathrm{L}0}$, $\Omega_0$, and $\gamma_0$,  respectively).
In practice, we observe that the boundaries for the generation of squeezing
depend only on the ratios $\Omega / \gamma$ and $\delta_\mathrm{L}/\gamma$
[see Eq.~\eqref{Eq5variancedephasing}].
For a QE-GNS configuration, these limits are shown in Fig.~\ref{fig2}a as a function of the rescaled detuning and driving field ($ \delta_0=2 \delta_\mathrm{L0}/\gamma_0$ and $z_0=\sqrt{2}\Omega_0/\gamma_0$, respectively).
Importantly, we find that
the detuning range with sizable squeezing has increased by two
orders of magnitude with respect to the case in free space, as displayed in
Fig.~\ref{fig2}b.
This is directly related to the enhancement of the QE spontaneous decay
rate $\gamma/\gamma_0\sim 60$ by the
GNS, which also leads to a shift in the resonance frequency.
Moreover, the nanostructure strongly influences the local field intensity at the position of the QE,
so that squeezing occurs over a much wider range of laser intensities
as compared to free space, \emph{c.f.}  Fig.~\ref{fig2}b at zero detuning.
The reason is that the GNS has a larger impact on the QE decay rate
than on the driving field enhancement with respect to
free space ($\Omega/ \Omega_0 \sim 4.9$ for this case), so that the
ratio $\Omega/\gamma$ provides a weaker excitation level at the same
incident power ($\propto z_{0} \Omega /\gamma$).

\begin{figure}
\begin{center}
\includegraphics[width=8.25cm,angle=0]{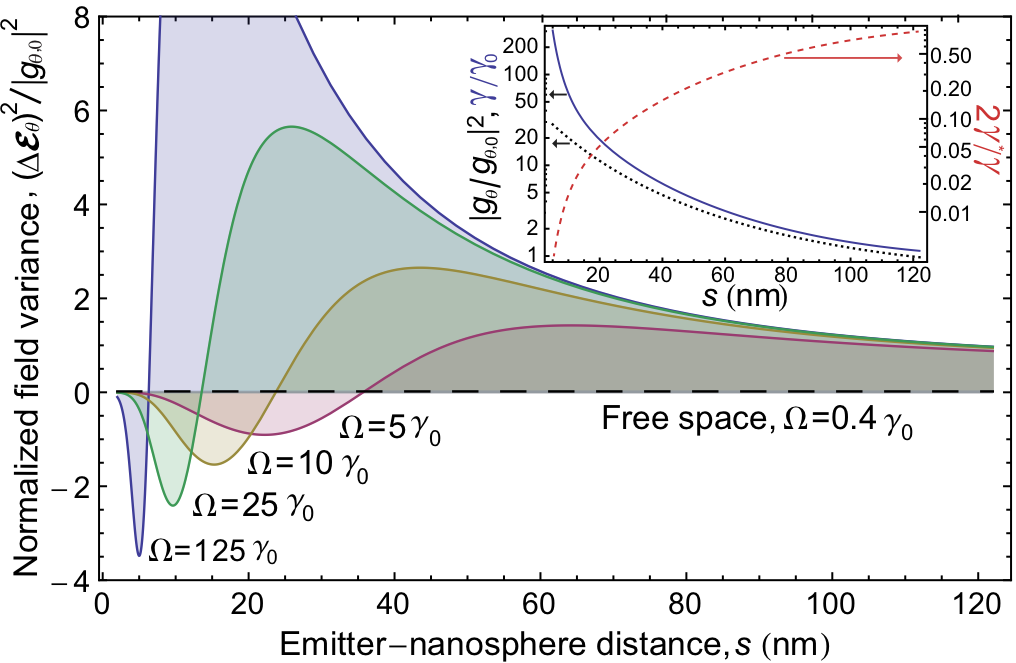}
\end{center}
\caption{(Color online) Normalized electric field fluctuations for the $\theta$-component as a function of the distance ${s}$ between the quantum emitter and the nanosphere
surface.
The quantum emitter is affected additionally by pure dephasing 
at a rate $\gamma^{*}=\gamma_{0}/2$.
The curves corresponds to Rabi frequencies $\Omega = 5 \gamma_{0}$, $10\gamma_0$,
$25\gamma_0$, and $125 \gamma_{0}$, respectively.
The other system parameters are the same as in Fig.~\ref{fig2}a.
For comparison, the result in the absence of the nanostructure and $\Omega=0.4\gamma_{0}$
is represented by the black dashed line.
The inset shows the normalized total decay rate $\gamma/\gamma_0$ (solid blue curve, left axis) and the field intensity enhancement factor
$|g_{\theta}/g_{\theta,0}|^2$ (dotted black curve, left axis) as a function of $s$.
The ratio ${2\gamma^{*}/\gamma}$ between the additional pure dephasing and the one
associated with spontaneous decay is also displayed (dashed red curve, right axis).
\label{fig3}}
\end{figure}

\paragraph*{Reduced quantum fluctuations under dephasing.}
Realistic QEs in free space are strongly affected by dephasing,\cite{batalov08} which can preclude the generation of squeezing.
To gain intuition on how the nanostructure may overcome this difficulty, we show  in Fig.~\ref{fig3} the
field fluctuations ${(\Delta\hat{\mathcal{E}}_{\theta})^{2}}$ as we vary the distance $s$
between the QE and the GNS surface (see Fig.~\ref{fig1}a) at zero detuning,
fixed Rabi frequency, and assuming an additional constant rate of pure dephasing, $\gamma^{*}=\gamma_{0}/2$.
In free space, ${(\Delta\hat{\mathcal{E}}_{\theta})^{2}}$
exhibits small positive values, \emph{i.e.} the field fluctuations
lie above shot noise.
In contrast, the presence of the GNS allows for quantum squeezing
over a range of distances $s$
that depend on the Rabi frequency, on the dephasing rate
and on the spontaneous decay.
For example, for $\Omega=5\gamma_0$, negative values of
${(\Delta\hat{\mathcal{E}}_{\theta})^{2}}$ occur
below \mbox{$s=35$~nm}  and its minimum is reached at \mbox{$s=23$~nm}.
This overall behavior is general, as highlighted by the other curves in Fig.~\ref{fig3}
corresponding to larger Rabi frequencies.
The minimum of each curve results from a balance between
the Rabi frequency, the decay rate $\gamma$, the ratio $2\gamma^{*}/\gamma$, and the amplitude $g_i$.
All of these depend on the emitter position (see the inset of Fig.~\ref{fig3})
while the Rabi frequency is kept constant.
Importantly, it is the large increase in the fluorescence rate $\gamma$
with respect to the free-space rates $\gamma^*$ and $\gamma_0$ that helps to fulfill the condition for squeezing in Eq.~\eqref{Eq5variancedephasing}.\footnote{The condition for squeezing, \emph{i.e.}
$(\Delta\hat{\mathcal{E}}_{i}(\mathbf{r},t))^{2}<0$
in Eq.~(\ref{Eq5variancedephasing}), sets
an upper limit for the driving field intensity, ${z^2<(1+\delta^2)(1-x)/(1+x)}$.
Notice that this requirement cannot be fulfilled for $x \ge 1$,
which already occurs for pure dephasing larger that half the spontaneous decay rate.}
As the QE moves towards the GNS surface, optimal squeezing requires increasingly stronger driving fields, especially once the distance $s$ falls below
10 nm, where absorption by real metals provides a dominating nonradiative decay channel for the QE.\cite{rogobete07a}
This is reflected in the growing deviation of $\gamma/\gamma_0$ from the
normalized radiative amplitude of squeezing $|g_{\theta}/g_{\theta,0}|^2$
(see inset in Fig.~\ref{fig3}). Nevertheless, the ratio between radiative and
nonradiative decay can be modified by optimized nanostructures\cite{rogobete07a}
and quantum squeezing may, in principle, be enhanced without considerably
raising the driving strengths to compensate for the nonradiative losses.
Thus the coupling of a QE to a nanostructure may facilitate the creation of
squeezed states of light despite of decoherence.

\begin{figure}
\begin{center}
\includegraphics[width=8.25cm,angle=0]{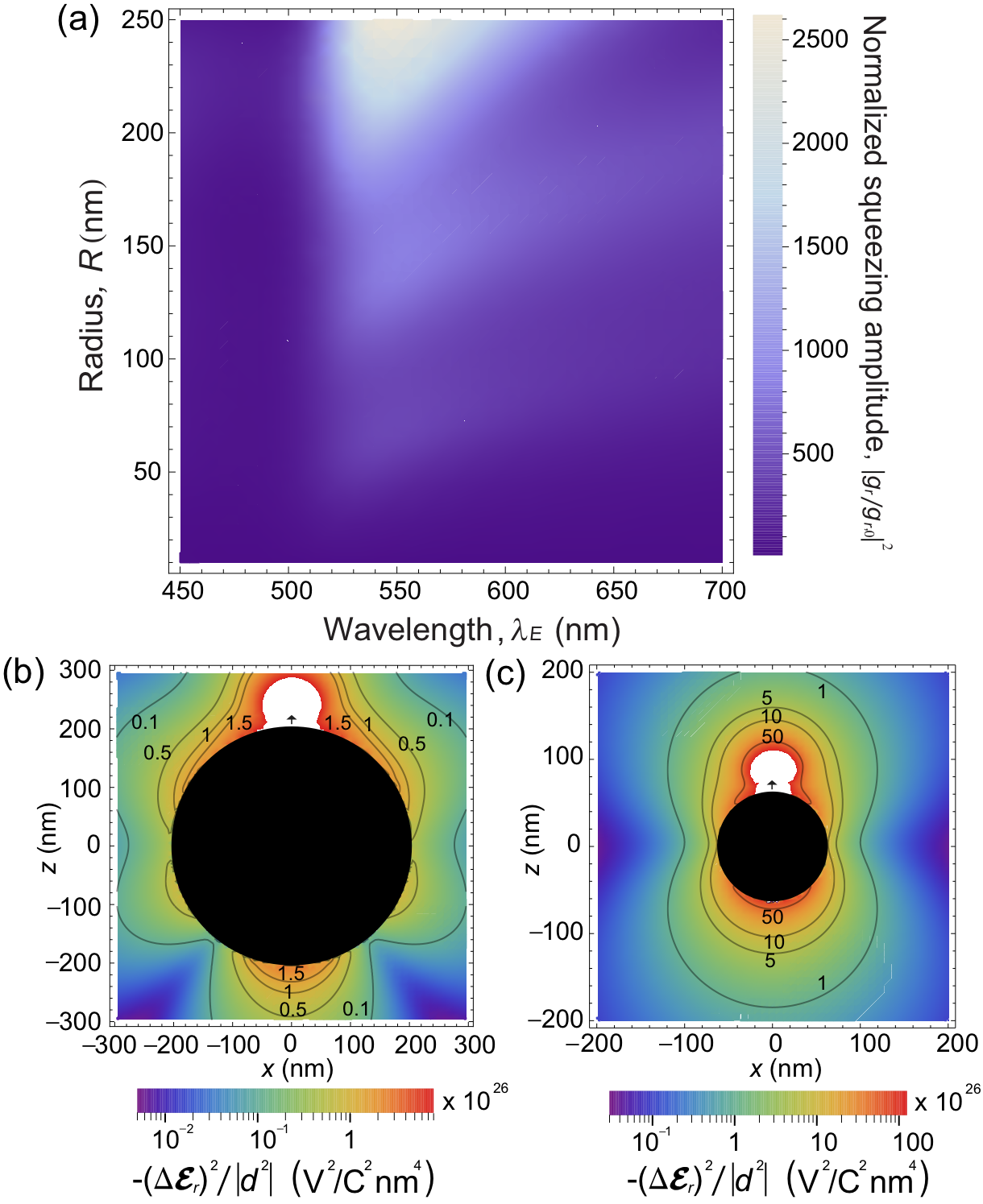}
\end{center}
\caption{(Color online) (a) Normalized squeezing amplitude for the radial near-field component
$|g_{r}/g_{r,0}|^{2}$ as a function of $\lambda_\mathrm{E}$ and $R$,
for an emitter-surface distance $s=10$ nm and detection at $D_2$ (see Fig.~\ref{fig1}a).
(b-c) Contour maps of the negative field fluctuations for the radial component for
$R=200$ nm (b) and $R=60$ nm (c), with $s=10$ nm and $\lambda_\mathrm{E}=550$ nm.
The values are normalized to the square modulus of the dipole moment $|\mathbf{d}|^2$ to be independent
of a specific quantum emitter.
The emitter and the nanosphere are represented by a black
arrow and a disk in the $xz$-plane, respectively.
\label{fig4}}
\end{figure}

\paragraph*{Reduced quantum fluctuations in the near field.}
Further enhancement of squeezing can be achieved in the near field, where intense evanescent modes
become relevant.
We emphasize that even in free space, the squeezing amplitude close to a QE is orders of magnitude higher as compared to the far field, due to the spatial behavior of its dipolar field.\cite{jackson99}
To estimate the ability of nanostructures to transport squeezed light away from the QE,
we consider a detection point on the opposite side of the GNS ($D_2$ in Fig.~\ref{fig1}a).
Figure~\ref{fig4}a displays the normalized squeezing amplitude for the
radial field component, $|g_{r}/g_{r,0}|^{2}$, which dominates
in the near-field region.
The enhancement leads to values two orders of
magnitude larger than those encountered in the far field
in the same parameter space explored by varying the wavelength
$\lambda_\mathrm{E}$ and the radius $R$.
In contrast with the far-field, the enhancement of squeezing increases with higher-order plasmon-polariton resonances at larger radii.
This fact arises from the rapid spatial decay of the radial field in free space
combined with the field enhancement near the GNS surface,
which boosts the ratio $|g_{r}/g_{r,0}|^2$.
Intuitively, we expect this quantity to increase up to very high radii until
the system resembles a QE near a flat metal
surface, where it becomes limited by propagation losses over the system size.\cite{dung02}

For a better understanding of the strong spatial dependence of the
squeezing amplitude in the near field,
we now analyze the electric field fluctuations over a cross-section of the
surrounding of the GNS.
Figure~\ref{fig4}b gives the near-field squeezing pattern
for a large GNS ($R=200$ nm).
We observe two lateral lobes, which stem from the excitation
of higher-order plasmon-polariton resonances.
Such squeezed field modes are superimposed with the dipolar contribution indicated by the presence of the top and bottom lobes, which are more clearly visible in Fig.~\ref{fig4}c in the case of a smaller GNS ($R=60$ nm), for which the dipole resonance prevails.
Note that despite the huge enhancements found for large GNSs compared to free space (see Fig.~\ref{fig4}a), the small GNS improves the squeezing amplitude, \emph{e.g.}, by a factor 30.
This is the result of a shorter detection distance with respect to the QE combined with a higher near-field enhancement.

\section{Conclusions}

Our study indicates a wide range of possibilities for
controlling the quantum fluctuations of light at the nanoscale
using a laser-driven QE coupled to a nanoarchitecture.
We found that the nanostructure-assisted dynamics of a QE improves the generation of squeezed light in resonance fluorescence, overcoming the limitations of weak driving.
An antenna effect\cite{kuehn06a} allows for boosting the
transfer of squeezing to the far field,
resulting in a large suppression of quantum fluctuations.
The huge enhancement of spontaneous decay
made possible by optical nanostructures\cite{rogobete07a} may also allow for
the generation of squeezed states of light under conditions
where the system undergoes fast dephasing.
Altogether, these findings facilitate the detection
of quantum squeezing in resonance fluorescence
from a single emitter
within the possibilities of current experiments
and provide perspectives for its practical application.
For instance, the large near fields can generate quantum fields
on the nanoscale with squeezing levels that are orders of magnitude higher
than in the far field.
These could be efficiently transferred over
a considerable distance by nanoscale waveguides.\cite{huck09}
Furthermore, since our approach can be applied to many different types of quantum emitters and nanostructures,
it may help to develop novel
solid-state sources of squeezed light for integrated nanophotonic
systems\cite{schietinger09,huck11,akimov07} and quantum-limited sensitivity,\cite{loudon87,taylor13}
and provide new insights into the production of
multi-partite entangled states.\cite{gonzalez-tudela11,gullans12}

\begin{acknowledgements}
Financial support from the Max Planck Society,
the EU-STREP project ``QIBEC'' and the MIUR-PRIN grant (2010LLKJBX) are gratefully acknowledged. M.A. wishes to thank Vahid Sandoghdar.
\end{acknowledgements}

\begin{widetext}

\appendix
\section{Electric field operator for the composite system}
\label{A1}

We use the macroscopic quantum electrodynamics formalism of Kn\"oll
\emph{et al.}~[\onlinecite{knoll01}] to write the
operator $\mathbf{\hat{E}}^{(+)}(\mathbf{r},t)$
of the positive-frequency electric field
scattered by a single quantum emitter, modeled as a two-level system (TLS),
coupled to a nanostructure of arbitrary shape. These electric field
operators can be used to obtain the correlations that identify
the generation of squeezed light, which are derived in Appendix~\ref{A2}.
Within the rotating wave approximation, the electric field operator
is written in terms of the
electromagnetic Green's tensor $\mathbf{G}$ and the emitter lowering operator
$\mathbf{\hat{\sigma}}$ as
\begin{equation}
\begin{split}
\label{Efield operatorHeisembergSup}
\mathbf{\hat{E}}^{(+)}_\mathrm{tot}(\mathbf{r},t)=&
\mathbf{\hat{E}}^{(+)}_{\textrm{free}}(\mathbf{r},t)+
\mathbf{\hat{E}}^{(+)}(\mathbf{r},t),\\
=&\mathbf{\hat{E}}^{(+)}_{\textrm{free}}(\mathbf{r},t)+
\frac{\imath}{\pi\varepsilon_{0}}\int^{\infty}_{0}\textrm{d}\omega\frac{\omega^{2}}{c^{2}}
\textrm{Im}\{\mathbf{G}(\mathbf{r},\mathbf{r}_\mathrm{E},\omega)\}\cdot\mathbf{d}\int^{t}_{t'}\textrm{d}\tau\,e^{-\imath\omega(t-\tau)}\mathbf{\hat{\sigma}}(\tau),
\end{split}
\end{equation}
where the first term represents the freely evolving part of the driving
electric field and the second term corresponds
to the source part scattered by the composite system. The quantum emitter
is located at $\mathbf{r}_\mathrm{E}$ and is characterized by its emission
frequency $\omega_E$ and its transition dipole matrix element $\mathbf{d}$.
The Markov approximation, in which $t' \to -\infty$, holds for
intervals $t-t'$ larger than the short correlation times in the presence
of the nanostructure, so that
the electric field operator can be expressed as\cite{knoll01}
\begin{equation}
\label{Efield operatorHeisembergSupMarkovRotatingaproxSup}
\mathbf{\hat{E}}^{(+)}_\mathrm{tot}(\mathbf{r},t)=
\mathbf{\hat{E}}^{(+)}_{\textrm{free}}(\mathbf{r},t)+\imath\hbar\mathbf{\hat{\sigma}}(t)\left(\boldsymbol{\gamma}(\mathbf{r})/2+
\imath\boldsymbol{\delta\omega}(\mathbf{r})\right),
\end{equation}
where
\begin{equation}
\label{coefficients}
\boldsymbol{\delta\omega}(\mathbf{r})=\frac{\mathcal{P}}{\pi\hbar\varepsilon_{0}}\int_{0}^{\infty}\textrm{d}\omega\,\frac{\omega^{2}}{c^{2}}
\frac{\textrm{Im}\{\mathbf{G}(\mathbf{r},\mathbf{r}_\mathrm{E},\omega)\}
\cdot\mathbf{d}}{\omega_\mathrm{E}-\omega},
\end{equation}
and
\begin{equation}
\label{gammaijbis}
\boldsymbol{\gamma}(\mathbf{r})=\frac{2\omega_\mathrm{E}^{2}}{\hbar\varepsilon_{0}c^{2}}\textrm{Im}\{\mathbf{G}(\mathbf{r},\mathbf{r}_\mathrm{E},\omega_\mathrm{E})\}
\cdot\mathbf{d},
\end{equation}
are vectors and $\omega_\mathrm{E}$ is the emitter resonance frequency.

We rewrite the source part of
Eq.~(\ref{Efield operatorHeisembergSupMarkovRotatingaproxSup})
as a complex vector, the $i$th-component of which is proportional to the
emitter lowering operator
\begin{equation}
\label{factorized_electric_fieldSup}
\hat{E}_{i}^{(+)}(\mathbf{r},t)=
\imath\hbar\hat{\sigma}(t)
\left(\gamma_{i}\mathbf{(r)}/2+\imath\delta\omega_{i}\mathbf{(r)}\right)=
|g_i(\mathbf{r})|e^{\imath\phi_i}\hat{\sigma}(t).
\end{equation}
The amplitude corresponds to
\begin{equation}
\label{mod_giSup}
|g_i(\mathbf{r})|=\hbar\sqrt{(\gamma_{i}\mathbf{(r)}/2)^{2}+(\delta\omega_{i}\mathbf{(r)})^{2}},
\end{equation}
and the complex phase is
\begin{equation}
\label{phase_giSup}
\phi_i(\mathbf{r})=\arctan\left(-\frac{\gamma_{i}\mathbf{(r)}}
{2 \delta\omega_{i}\mathbf{(r)}}\right).
\end{equation}

In a general geometry, the amplitude involved in the source part of
Eq.~(\ref{Efield operatorHeisembergSupMarkovRotatingaproxSup})
has to be evaluated numerically. A semi-analytical treatment of the
Green's tensor is possible for the case of a gold nanosphere (GNS)
considered in this work.\cite{dung01,li94}
In particular, the frequency integral in Eq.~(\ref{coefficients})
can be calculated in the complex plane,
where the tabulated optical constants of gold\cite{CRChandbook} are
replaced by a Drude-Lorentz dispersion model.\cite{kaminski07} 

\begin{figure}
\begin{center}
\includegraphics[width=17cm,angle=0]{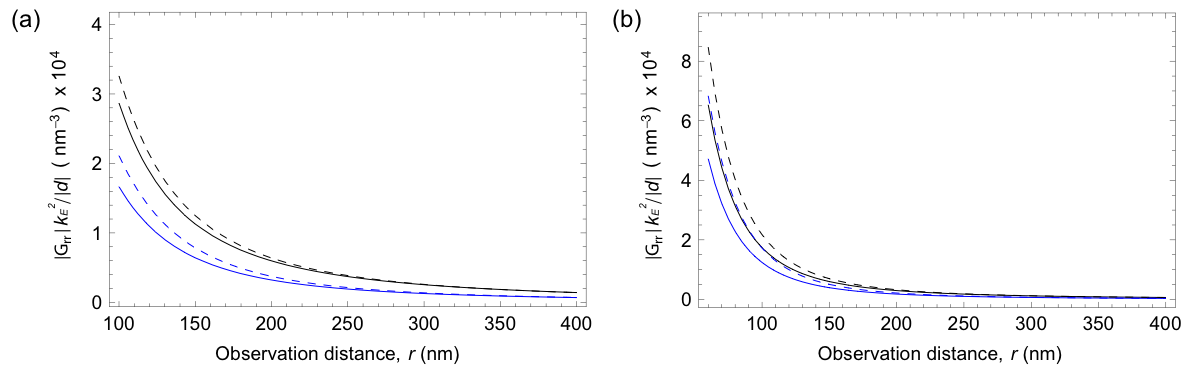}
\end{center}
\caption{Absolute value of the radial-radial component of the Green's tensor
(dashed curves) and the full contribution to the radial component
of the amplitude $|g_{r}|$ in
Eq.~(\ref{Efield operatorHeisembergSupMarkovRotatingaproxSup})
(solid curves)
as a function of the observation distance $r$ for $R=80$ nm (a)
and $R=45$ nm (b), and two different emission wavelengths, $\lambda=600$ nm
(black curves) and $\lambda=800$ nm (blue curves), respectively.
Their values are multiplied by the emission wavevector $k_\mathrm{E}^{2}/|\mathbf{d}|$
which corresponds to units of nm$^{-3}$.
The emitter separation from the GNS corresponds to $s=10$ nm, observation
angle $\theta=\pi/2$ (see Fig.~\ref{fig1}) and the emitter
dipole-moment is perpendicular to the GNS surface.
\label{fig1NearfieldcomparisonSupp}}
\end{figure}

The principal value integral can be simplified by neglecting off-resonant
contributions in the intermediate and far-field region
($|\mathbf{r}-\mathbf{r}_\mathrm{E}| \gg \lambda$).\cite{dung02b,dzsotjan11}
The impact of this approximation can be seen in
Fig.~\ref{fig1NearfieldcomparisonSupp}, where we compare the
absolute values for the radial-radial Green's tensor components
obtained from the exact integral
($I(\mathbf{r},\mathbf{r}_\mathrm{E},\omega_\mathrm{E}) =
|\mathcal{P}\int \textrm{d}\omega
[\textrm{Im}\{G_{rr}(\mathbf{r},\mathbf{r}_\mathrm{E},\omega)\}/(\omega_\mathrm{E}-\omega)]+
\imath\textrm{Im}\{G_{rr}(\mathbf{r},\mathbf{r}_\mathrm{E},\omega_\mathrm{E})\}|$, solid curves) and from
the approximated one
($I(\mathbf{r},\mathbf{r}_\mathrm{E},\omega_\mathrm{E}) \approx
|\textrm{Re}\{G_{rr}(\mathbf{r},\mathbf{r}_\mathrm{E},\omega_\mathrm{E})\}+
\imath\textrm{Im}\{G_{rr}(\mathbf{r},\mathbf{r}_\mathrm{E},\omega_\mathrm{E})\}|$,
dashed curves).
We have chosen two emission wavelengths and two different radii
($R=80$ nm in panel (a) and $R=45$ nm in panel (b)). For all curves
we observe how the difference between the exact and approximated
values increases for smaller distances and becomes negligible
for large ones. Moreover the deviation is larger in the case of smaller
GNSs,\cite{dung02b} as the comparison between panels (a) and (b)
shows.
In addition, the contribution of the off-resonant frequencies is more
important for increasing wavelengths.
For instance at $\lambda = 800$ nm (blue curves) and at an
observation distance of 20 nm from the GNS surface, it amounts to
$27\%$ and $42\%$ of $G_{rr}$ for $R=80$ nm and $R=45$ nm,
respectively. 
Hence, the full frequency integral has been
used in the analysis of the electromagnetic near fields in
the manuscript (see Fig.~\ref{fig4}),
otherwise the far-field approximation has been imposed for
numerical efficiency (see Figs.~\ref{fig1}-\ref{fig3}).

\section{Evaluation of the normally-ordered electric field variance}
\label{A2}

We now derive the expressions in Eqs.~(1) and (2).
Defining a general quadrature operator of the electric field as
$\mathbf{\hat{E}}_{\Theta}=e^{\imath\Theta}\mathbf{\hat{E}}^{(+)}+
e^{-\imath\Theta}\mathbf{\hat{E}}^{(-)}$, we write its variance
$\langle \mathbf{\hat{E}}_{\Theta}^2\rangle-\langle
\mathbf{\hat{E}}_{\Theta}\rangle^2$
in normal order ($\langle::\rangle$) as
\begin{equation}
\label{variance_quadratureSup}
\begin{split}
\mean{:\![\Delta\mathbf{\hat{E}}_{\Theta}(\mathbf{r},t)]^{2}\!:} = &
e^{\imath 2\Theta}\mean{[\mathbf{\hat{E}}^{(+)}(\mathbf{r},t)]^2}+
e^{-\imath 2\Theta}\mean{[\mathbf{\hat{E}}^{(-)}(\mathbf{r},t)]^2}-
2\textrm{Re}\{e^{\imath 2\Theta}
\mean{\mathbf{\hat{E}}^{(+)}(\mathbf{r},t)}^{2}\} \\
& +2\mean{\mathbf{\hat{E}}^{(-)}(\mathbf{r},t)
 \mathbf{\hat{E}}^{(+)}(\mathbf{r},t)}-
2\mean{\mathbf{\hat{E}}^{(-)}(\mathbf{r},t)}
\mean{\mathbf{\hat{E}}^{(+)}(\mathbf{r},t)},
\end{split}
\end{equation}
where $\mathbf{\hat{E}}^{(-)}=(\mathbf{\hat{E}}^{(+)})^{\dag}$
corresponds to the negative-frequency electric field.
Next we use Eqs.~(\ref{factorized_electric_fieldSup})-(\ref{phase_giSup})
in Eq.~(\ref{variance_quadratureSup}) to obtain
the normally-ordered variance for the $i$th-component of the electric field
quadrature
\begin{equation}
\label{variance_finalSup}
\mean{:\![\Delta \hat{E}_{i,\Theta}(\mathbf{r},t)]^{2}\!:}  =
2|g_i(\mathbf{r})|^2
\left[\left(\mean{\hat{\sigma}^{\dag}(t)\hat{\sigma}(t)}
-|\mathbf{\hat{\sigma}}(t)|^2\right)-\textrm{Re}\{e^{\imath 2(\Theta+\phi_i)}
\mean{\mathbf{\hat{\sigma}}(t)}^{2}\}\right]
 = |g_i(\mathbf{r})|^2 \mean{:\![\Delta \hat{\sigma}_{\Theta+\phi_{i}}]^{2}\!:},
\end{equation}
which for $\Theta=0$ gives Eq.~(1),
where we have simplified the notation by introducing the symbol ${(\Delta\hat{\mathcal{E}}_{i}(\mathbf{r},t))^{2}}$.
Here $\mean{:\![\Delta \hat{\sigma}_{\Theta+\phi_{i}}]^{2}\!:}$ denotes the general normally-ordered variance of the
TLS coherence quadrature operator,
$\hat{\sigma}_{\Theta}=e^{\imath\Theta}\hat{\sigma}+
e^{-\imath\Theta}\hat{\sigma}^{\dag}$,
evaluated at the angle $\Theta+\phi_{i}$.

The variance in Eq.~(\ref{variance_finalSup}) can be evaluated
starting from the expectation value of the slowly varying coherence
$\mean{\hat{\tilde{\sigma}}(t)}=\mean{\hat{\sigma}(t)}e^{\imath\omega_{L}t}$ in the co-rotating frame,
which is governed by modified optical Bloch equations\cite{knoll01,chen13}
\begin{eqnarray}
\label{Bloch_equations1}
\partial_t \mean{\hat{\tilde{\sigma}}(t)}&=&
\left(-\frac{\gamma}{2}-\gamma^{*}+\imath\delta_\mathrm{L}\right)
\mean{\hat{\tilde{\sigma}}(t)}-
\imath\frac{\Omega}{2}\mean{\hat{\sigma}_{z}(t)}, \\
\partial_t\mean{\hat{\sigma}_{z}(t)}&=& \imath\left(\Omega
\mean{\hat{\tilde{\sigma}}^{\dag}(t)}-\Omega^{*}
\mean{\hat{\tilde{\sigma}}(t)}\right)-
\gamma(1+\mean{\hat{\sigma}_{z}(t)}), \label{Bloch_equations2}
\end{eqnarray}
where $\Omega=2\mathbf{d}\cdot
\mean{\hat{\mathbf{E}}^{(+)}_{\textrm{free}}
(\mathbf{r}_\mathrm{E})}/\hbar=|\Omega|e^{\imath\phi_\mathrm{L}}$
is the Rabi frequency and includes the local driving field enhancement due to the GNS.\cite{mojarad08} The spontaneous decay rate is
$\gamma=\frac{2\omega_\mathrm{E}^{2}}{\hbar\varepsilon_{0}c^{2}}
\mathbf{d} \cdot \textrm{Im}\{\mathbf{G}(\mathbf{r}_\mathrm{E},
\mathbf{r}_\mathrm{E},\omega_\mathrm{E})\} \cdot\mathbf{d}$,
$\gamma^{*}$ is the pure
dephasing rate , and $\delta_\mathrm{L}=\omega_\mathrm{L}-
\tilde{\omega}_\mathrm{E}$ is the detuning with respect to
the dressed transition frequency
$\tilde{\omega}_\mathrm{E}=\omega_\mathrm{E}-
\frac{\mathcal{P}}{\pi\hbar\varepsilon_{0}}
\int_{0}^{\infty}\textrm{d}\omega\,[\frac{\omega^{2}}
{c^{2}}\mathbf{d}\cdot\textrm{Im}\{\mathbf{G}
(\mathbf{r}_\mathrm{E},\mathbf{r}_\mathrm{E},\omega)\}
\cdot\mathbf{d}/(\omega-\omega_\mathrm{E})]$.
From the steady-state condition of Eqs.~(\ref{Bloch_equations1}) and
(\ref{Bloch_equations2}), we deduce the stationary expectation
value of the slow varying coherence
$\mean{\hat{\tilde{\sigma}}}_\mathrm{s}$ and population
$\mean{\hat{\sigma}_z}_\mathrm{s}$
\begin{eqnarray}
\mean{\hat{\tilde{\sigma}}}_\mathrm{s}&=&\frac{-\Omega
[2\delta_\mathrm{L}-\imath ( \gamma +2 \gamma^{*}) ]}
{4\delta_\mathrm{L}^{2}+2|\Omega|^{2}(1+\frac{2\gamma^{*}}
{\gamma})+(\gamma+2\gamma^{*})^{2}}
 =e^{\imath[\phi_\mathrm{L}+\phi_\mathrm{dep}]}
\sqrt{\frac{1}{2(1+x)}}\frac{z(\sqrt{\delta^2+1})}
{1+\delta^2+z^2},\label{stationaryvalues}\\
\mean{\hat{\sigma}_z}_\mathrm{s}&=&-1+\imath\left(\frac{\Omega}{\gamma}
\mean{\hat{\tilde{\sigma}}^{\dag}(\infty)}-
\frac{\Omega^{*}}{\gamma}\mean{\hat{\tilde{\sigma}}(\infty)}\right)=-\frac{1+\delta^2}{1+\delta^2+z^2},\label{stationaryvalues2}
\end{eqnarray}
in which we have used the normalized Rabi frequency
${z=\sqrt{2} |\Omega|/\sqrt{\gamma (\gamma+2\gamma^{*})}}$,
the coherence dephasing phase $\phi_\mathrm{dep}=
\arctan\left[-(\gamma +2 \gamma^{*})/(2\delta_\mathrm{L})\right]$,
the normalized pure dephasing $x=2\gamma^{*}/\gamma$, and
the normalized detuning $\delta=2\delta_{L}/(\gamma+2\gamma^{*})$.
Notice that the normalized detuning and normalized Rabi frequency
can be written in terms of the free-space normalized
variables $\delta_0$ and $z_0$ as
$\delta=[\delta_0-2(\omega_\mathrm{E}-\tilde{\omega}_\mathrm{E})/
\gamma_0]/[(\gamma/\gamma_0)(1+x)]$
and
$z=z_0|\Omega|/[|\Omega_0|(\gamma/\gamma_0)\sqrt{(1+x)}]$, respectively,
where $\gamma_0$ is the decay rate and $\Omega_0$ the Rabi frequency
in free space. Using the results of Eqs.~(\ref{stationaryvalues}) and
(\ref{stationaryvalues2}) and the stationary expectation value of the coherence
$\mean{\hat{\sigma}}_\mathrm{s}=\mean{\hat{\tilde{\sigma}}}_\mathrm{s}
e^{-\imath \omega_{L}t}$,
we obtain the general normally-ordered atomic variance
\begin{equation}
\begin{split}
\label{variancedephasingSup}
\mean{:\![\Delta \hat{\sigma}_{\Theta}]^{2}\!:}=&
2\left(\mean{\hat{\sigma}^{\dag}\hat{\sigma}}_\mathrm{s}-
|\mean{\hat{\sigma}}_\mathrm{s}|^{2}\right)-
2\textrm{Re}\{e^{\imath 2\Theta}\mean{\hat{\sigma}}_\mathrm{s}^{2}\}\\
=&(1+\mean{\hat{\sigma}_{z}}_\mathrm{s})-2|\mean{\hat{\sigma}}_\mathrm{s}|^{2}-
2\textrm{Re}\{e^{\imath 2\Theta}\mean{\hat{\sigma}}_\mathrm{s}^{2}\}\\
=&\frac{z^2}{1+\delta^2+z^2}\left\{1-
\frac{1}{1+x}\frac{(\delta^2+1)
[1+\cos(2\Theta+2\phi_\mathrm{L}+
2\phi_\mathrm{dep}-2\omega_\mathrm{L}t)]}
{1+\delta^2+z^2}\right\},
\end{split}
\end{equation}
in which we have used the general property
[$\mean{\hat{\sigma}^{\dag}\hat{\sigma}}=
(1+\mean{\hat{\sigma}_z})/2$] for a TLS,
that is derived from the commutators
$[\hat{\sigma}^{\dag},\hat{\sigma}]=\hat{\sigma}_{z}$
and the anti-commutation relation
$\{\hat{\sigma}^{\dag},\hat{\sigma}\}=\mathbb{I}$.
In general, $\Theta$ represents the phase of the electric field
quadrature plus any propagative phase of $g_i$,
$\phi_\mathrm{L}$ is the phase of the driving field at $\mathbf{r}_\mathrm{E}$,
which is affected by the GNS, and $\phi_\mathrm{dep}$ is the phase of the
steady-state coherence $\langle\hat{\sigma}\rangle_\mathrm{s}$.
Equation~(\ref{variance_finalSup}) for $\Theta=0$ together with
Eq.~(\ref{variancedephasingSup}) for $\Theta=\phi_i$ give the desired
Eq.~(2), where we denote
$\mean{:\![\Delta \hat{E}_{i,0}(\mathbf{r},t)]^{2}\!:} \equiv {(\Delta\hat{\mathcal{E}}_{i}(\mathbf{r},t))^{2}}$ for brevity.
Squeezing occurs if Eq.~(\ref{variancedephasingSup}) takes negative values.
For the optimal condition when the cosine is equal to 1, 
this leads to the following threshold for the driving intensity
\begin{equation}
z^2 < (1 + \delta^2) \frac{1-x}{1+x},
\end{equation}
which is discussed in the main text.

\section{Electric field variance from detection measurements}
\label{A3}

The measurement of squeezed light is associated with the photon
correlations that are obtained from the counting statistics of photodetectors.
Balanced homodyne detection is a well-established experimental
procedure for measuring the correlation
$\mean{:\![\Delta\mathbf{\hat{E}}_{\Theta}(\mathbf{r},t)]^{2}\!:}$.\cite{ourjoumtsev11,mandel82,vogel91,collett84}
Typically, in the homodyne detection scheme, the source field in
Eq.~(\ref{Efield operatorHeisembergSupMarkovRotatingaproxSup}) is mixed
by a beam splitter with a local oscillator, namely a high-intensity
coherent field.
 This is a laser field $\mathbf{\hat{E}}_{\textrm{free}}^{(+)}(\mathbf{r},t)\ket{\alpha}=
|\alpha| e^{\imath(\phi_\mathrm{L}-\omega_\mathrm{L}t)}\ket{\alpha}$,
where $\ket{\alpha}$ is a coherent state of amplitude $|\alpha|$
and well-defined phase $\phi_\mathrm{L}$ at the photodetector, so that fluctuations are due to shot noise, \emph{i.e.},
$\mean{:\!\Delta\mathbf{\hat{E}}_{\textrm{free}}\!:}=0$.
Note that in the setup sketched in Fig.~\ref{fig1}a,
the total field is actually a superposition of
$\mathbf{\hat{E}}_{\textrm{free}}(\mathbf{r},t)$
and the source field $\mathbf{\hat{E}}(\mathbf{r},t)$, so that it
already contains the local oscillator and does not
require a beam splitter.
Assuming that the coherent field is much stronger than the source field
$\mathbf{\hat{E}}_{\textrm{free}}(\mathbf{r},t)\gg
\mathbf{\hat{E}}(\mathbf{r},t)$, one can derive
the variance of the photocounts $\overline{\Delta n}$ for a
detector within the short time interval $\Delta t$
\begin{equation}
\label{variance_counts}
\overline{\Delta n}^2=\xi \Delta t |\alpha|^2 +
\xi^2 \Delta t^2 |\alpha|^2
\mean{:\![\Delta\mathbf{\hat{E}}_{-\phi_\mathrm{L}+
\omega_\mathrm{L}t}(\mathbf{r},t)]^{2}\!:},
\end{equation}
where $\xi$ is the detector efficiency.
From Eq.~(\ref{variance_counts}), it is inferred that
\mbox{$\mean{:\!\Delta[\mathbf{\hat{E}}_{-\phi_\mathrm{L}+
\omega_\mathrm{L}t}(\mathbf{r},t)]^{2}\!:}$} is obtained
 by subtracting and normalizing the detected shot noise $\overline{n}=\xi \Delta t |\alpha|^2$, which provides (see the derivation in Appendix~\ref{A3.1})
\begin{equation}
\label{variance_counts_vs_shot}
\frac{\overline{\Delta n}^2-\overline{n}}
{ \overline{n}}=\xi \Delta t \mean{:\![\Delta
\mathbf{\hat{E}}_{-\phi_\mathrm{L}+\omega_\mathrm{L}t}
(\mathbf{r},t)]^{2}\!:}.
\end{equation}
The scattered field is squeezed if this magnitude has a
negative value, \emph{i.e.}, it features reduced quantum
fluctuations with respect to shot noise, which is proportional to $|g_i(\mathbf{r})|^2$ as inferred from Eq.~(\ref{variance_finalSup}). Moreover,
the phase of the laser can be varied in the far field in order to fix the
phase in Eq.~(\ref{variancedephasingSup}) and maximize the
cosine, \emph{i.e.}, maximize the degree of squeezing.

\subsection{Derivation of Equation~(\ref{variance_counts_vs_shot})}
\label{A3.1}

In order to derive Eq.~(\ref{variance_counts_vs_shot})
we need the mean number of photocounts in the time interval $\Delta t$.
Using a detector with efficiency $\xi$ this quantity reads\cite{vogel06}
\begin{equation}
\label{mean_counts}
\begin{split}
\overline{n}= & \xi \Delta t
\mean{\mathbf{\hat{E}}^{(-)}_\mathrm{tot}(\mathbf{r},t)
\mathbf{\hat{E}}^{(+)}_\mathrm{tot}(\mathbf{r},t)},\\
= &
\xi \Delta t \left( \mean{\mathbf{\hat{E}}_{\textrm{free}}^{(-)}
(\mathbf{r},t)\mathbf{\hat{E}}_{\textrm{free}}^{(+)}(\mathbf{r},t)}
+
\mean{\mathbf{\hat{E}}^{(-)}(\mathbf{r},t)
\mathbf{\hat{E}}^{(+)}(\mathbf{r},t)}
+
\mean{\mathbf{\hat{E}}_{\textrm{free}}^{(-)}(\mathbf{r},t)
\mathbf{\hat{E}}^{(+)}(\mathbf{r},t)}+
\mean{\mathbf{\hat{E}}^{(-)}(\mathbf{r},t)
\mathbf{\hat{E}}_{\textrm{free}}^{(+)}(\mathbf{r},t)} \right).
\end{split}
\end{equation}
Here, we used the free and source field notation of
Eq.~(\ref{Efield operatorHeisembergSupMarkovRotatingaproxSup}).
If we introduce the density matrix
$\hat{\rho}=\hat{\sigma}\ket{\alpha}\!\bra{\alpha}$ to describe a disentangled
emitter-field system at equal times, Eq.~(\ref{mean_counts}) becomes
\begin{equation}
\label{mean_counts_laser_disentangled}
\overline{n}=\xi \Delta t \left( |\alpha|^2+
\mean{\mathbf{\hat{E}}^{(-)}\mathbf{\hat{E}}^{(+)}}+
|\alpha| e^{-\imath(\phi_\mathrm{L}+\omega_\mathrm{L}t)}
\mean{\mathbf{\hat{E}}^{(+)}}+ |\alpha| e^{\imath (\phi_\mathrm{L}-\omega_\mathrm{L}t)}
\mean{\mathbf{\hat{E}}^{(-)}} \right).
\end{equation}
If we further assume that $|\alpha|^2\gg
\mean{\mathbf{\hat{E}}^{(-)}\mathbf{\hat{E}}^{(+)}}$,
the number of photocounts is mainly determined by the laser field
\begin{equation}
\label{laser counts}
\overline{n}=\xi \Delta t  |\alpha|^2.
\end{equation}
Moreover, we need
the variance in the number of photocounts, which follows from statistical considerations\cite{vogel06}
\begin{equation}
\label{variance_counts_general}
\overline{\Delta n}^2=\overline{n}+\xi^2 (\Delta t)^{2}
\mean{:\![\Delta\mathbf{\hat{I}}(\mathbf{r},t)]^{2}\!:},
\end{equation}
where $\mathbf{\hat{I}}(\mathbf{r},t)=
\mathbf{\hat{E}}_\mathrm{tot}^{(-)}(\mathbf{r},t)
\mathbf{\hat{E}}_\mathrm{tot}^{(+)}(\mathbf{r},t)$
is the first-order intensity correlation.
To evaluate Eq.~(\ref{variance_counts_general}) we write the intensity variance  explicitly
\begin{equation}
\label{normalvariance_intensity}
\Delta \mathbf{\hat{I}}(\mathbf{r},t) =  \Delta\{
\mathbf{\hat{E}}^{(-)}(\mathbf{r},t)\mathbf{\hat{E}}^{(+)}(\mathbf{r},t) \}
+\Delta \{\mathbf{\hat{E}}_{\textrm{free}}^{(-)}(\mathbf{r},t)
\mathbf{\hat{E}}_{\textrm{free}}^{(+)}(\mathbf{r},t) \}+
\Delta \{\mathbf{\hat{E}}^{(-)}(\mathbf{r},t)
\mathbf{\hat{E}}_{\textrm{free}}^{(+)}(\mathbf{r},t)+
\mathbf{\hat{E}}_{\textrm{free}}^{(-)}(\mathbf{r},t)
\mathbf{\hat{E}}^{(+)}(\mathbf{r},t)  \},
\end{equation}
and
\begin{equation}
\label{squarednormalvariance_intensity}
\begin{split}
[\Delta \mathbf{\hat{I}}(\mathbf{r},t)]^{2}= &
[\Delta\{\mathbf{\hat{E}}^{(-)}\mathbf{\hat{E}}^{(+)} \}]^{2}+
[\Delta \{\mathbf{\hat{E}}_{\textrm{free}}^{(-)}\mathbf{\hat{E}}_{\textrm{free}}^{(+)} \}]^{2}+
[\Delta \{\mathbf{\hat{E}}^{(-)}\mathbf{\hat{E}}_{\textrm{free}}^{(+)}+\mathbf{\hat{E}}_{\textrm{free}}^{(-)}\mathbf{\hat{E}}^{(+)}  \}]^{2}+\\
&
\Delta\{\mathbf{\hat{E}}^{(-)}\mathbf{\hat{E}}^{(+)}\} \Delta \{\mathbf{\hat{E}}_{\textrm{free}}^{(-)}\mathbf{\hat{E}}_{\textrm{free}}^{(+)}\}+
\Delta \{\mathbf{\hat{E}}_{\textrm{free}}^{(-)}\mathbf{\hat{E}}_{\textrm{free}}^{(+)}\} \Delta\{\mathbf{\hat{E}}^{(-)}\mathbf{\hat{E}}^{(+)}\}+\\
&
\Delta\{\mathbf{\hat{E}}^{(-)}\mathbf{\hat{E}}^{(+)}\}\Delta \{\mathbf{\hat{E}}^{(-)}\mathbf{\hat{E}}_{\textrm{free}}^{(+)}+\mathbf{\hat{E}}_{\textrm{free}}^{(-)}\mathbf{\hat{E}}^{(+)}\}+
\Delta \{\mathbf{\hat{E}}^{(-)}\mathbf{\hat{E}}_{\textrm{free}}^{(+)}+\mathbf{\hat{E}}_{\textrm{free}}^{(-)}\mathbf{\hat{E}}^{(+)}\}\Delta\{\mathbf{\hat{E}}^{(-)}\mathbf{\hat{E}}^{(+)}\}+\\
&
\Delta \{\mathbf{\hat{E}}_{\textrm{free}}^{(-)}\mathbf{\hat{E}}_{\textrm{free}}^{(+)}\}\Delta \{\mathbf{\hat{E}}^{(-)}\mathbf{\hat{E}}_{\textrm{free}}^{(+)}+\mathbf{\hat{E}}_{\textrm{free}}^{(-)}\mathbf{\hat{E}}^{(+)}\}+
\Delta \{\mathbf{\hat{E}}^{(-)}\mathbf{\hat{E}}_{\textrm{free}}^{(+)}+\mathbf{\hat{E}}_{\textrm{free}}^{(-)}\mathbf{\hat{E}}^{(+)}\}\Delta\{\mathbf{\hat{E}}_{\textrm{free}}^{(-)}\mathbf{\hat{E}}_{\textrm{free}}^{(+)}\}.
\end{split}
\end{equation}
Introducing this last expression in Eq.~(\ref{variance_counts_general})
and using the density matrix
$\hat{\rho}=\hat{\sigma}\ket{\alpha}\!\bra{\alpha}$, we find the expectation value
\begin{equation}
\begin{split}
\label{variance_countsfull_general}
\overline{\Delta n}^2=&
\overline{n}+ \xi^2 \Delta t^2 \left(
\mean{:\![\Delta\{\mathbf{\hat{E}}^{(-)}\mathbf{\hat{E}}^{(+)} \}]^{2}\!:}+
|\alpha|^2\mean{:\![\Delta \{\mathbf{\hat{E}}^{(-)}
e^{\imath(\phi_\mathrm{L}-\omega_\mathrm{L}t)}+e^{-\imath(\phi_\mathrm{L}+\omega_\mathrm{L}t)}
\mathbf{\hat{E}}^{(+)} \}]^{2}\!:} +\right.\\
&\left. 2 |\alpha|\mean{:\!\Delta\{\mathbf{\hat{E}}^{(-)}
\mathbf{\hat{E}}^{(+)}\}\Delta
\{\mathbf{\hat{E}}^{(-)} e^{\imath(\phi_\mathrm{L}-\omega_\mathrm{L}t)}+
e^{-\imath(\phi_\mathrm{L}+\omega_\mathrm{L}t)}\mathbf{\hat{E}}^{(+)} \}\!:} \right),
\end{split}
\end{equation}
where we have used the fact that the correlations
$\mean{:\!\Delta \{\mathbf{\hat{E}}_{\textrm{free}}^{(-)}\mathbf{\hat{E}}_{\textrm{free}}^{(+)}\}\!:}=0$
due to the coherent character of the free field.
If we assume the same conditions considered for deriving
Eq.~(\ref{laser counts}), then the third term
in Eq.~(\ref{variance_countsfull_general}) dominates and we obtain
\begin{equation}
\label{variance_counts_vs_shotfull}
\overline{\Delta n}^2=
\overline{n}+\xi^{2} (\Delta t)^{2} |\alpha|^2\mean{:\![\Delta
\{\mathbf{\hat{E}}^{(-)} e^{\imath(\phi_\mathrm{L}-\omega_\mathrm{L}t)}+
e^{-\imath(\phi_\mathrm{L}+\omega_\mathrm{L}t)}
\mathbf{\hat{E}}^{(+)} \}]^{2}\!:},
\end{equation}
from which we deduce the correlation formula
of Eq.~(\ref{variance_counts_vs_shot}).

\end{widetext}

%

\end{document}